\documentclass[aps,pre,onecolumn,superscriptaddress,notitlepage]{revtex4-1}
\usepackage[utf8]{inputenc}
\usepackage[T1]{fontenc}
\usepackage{lmodern}
\usepackage{amsmath,amssymb}
\pdfoutput=1
\usepackage{xcolor}
\usepackage{times}
\usepackage{graphicx}
\usepackage{hyperref}
\usepackage{tikz}

\usepackage{amsmath,amssymb}
\usepackage{xcolor}
\usepackage{times}
\usepackage{graphicx}

\newcommand*\colvec[3][]{
    \begin{pmatrix}\ifx\relax#1\relax\else#1\\\fi#2\\#3\end{pmatrix}
}
\renewcommand{\vec}[1]{\mathbf{#1}}

\newcommand\cP{{\cal P}}

\usepackage{hyperref}
\renewcommand\r{{\bf r}}
\begin{document} 
\title{Pressure is not a state function for generic active fluids}
\author{Alexandre P. Solon}
\affiliation{Universit\'e Paris Diderot, Sorbonne Paris Cit\'e, MSC, UMR 7057 CNRS, 75205 Paris, France}
\author{Y. Fily}
\affiliation{Martin Fisher School of Physics, Brandeis University, Waltham, MA 02453, USA}
\author{A. Baskaran}
\affiliation{Martin Fisher School of Physics, Brandeis University, Waltham, MA 02453, USA}
\author{M. E. Cates}
\affiliation{SUPA, School of Physics and Astronomy, University of Edinburgh, JCMB, Peter Guthrie Tait Road, Edinburgh EH9 3FD, UK}
\author{Y. Kafri}
\affiliation{Department of Physics, Technion, Haifa 32000, Israel}
\author{M. Kardar}
\affiliation{Department of Physics, Massachusetts Institute of Technology, Cambridge, Massachusetts 02139, USA}
\author{J. Tailleur}
\affiliation{Universit\'e Paris Diderot, Sorbonne Paris Cit\'e, MSC, UMR 7057 CNRS, 75205 Paris, France}
\date{\today}

\begin{abstract}
  Pressure is the mechanical force per unit area that a confined system
  exerts on its container. In thermal equilibrium, it depends only on
  bulk properties (density, temperature, etc.) through an equation of
  state. Here we show that in a wide class of active systems the
  pressure depends on the precise interactions between the active
  particles and the confining walls. In general, therefore, active
  fluids have no equation of state; their mechanical pressures exhibit
  anomalous properties that defy the familiar thermodynamic reasoning
  that holds in equilibrium. The pressure remains a function of state,
  however, in some specific and well-studied active models that tacitly
  restrict the character of the particle-wall and/or particle-particle
  interactions.
\end{abstract}

\maketitle 

For fluids in thermal equilibrium, the concept of pressure, $P$, is
familiar as the force per unit area exerted by the fluid on its
containing vessel. This primary, mechanical definition of pressure
{\em seems} to require knowledge of the interactions between the
fluid's constituent particles and its confining walls.  But we learn
from statistical mechanics that $P$ can also be expressed
thermodynamically, as the derivative of a free energy with respect to
volume. The pressure therefore obeys an equation of state, which only
involves bulk properties of the fluid (temperature $T$, number density
$\rho$, etc.).   Hydrodynamics provides a third definition of
$P$, as the trace of the bulk thermodynamic stress tensor, whose
microscopic definition in terms of momentum fluxes is again well
known~\cite{Allen}. In thermal equilibrium, all these definitions of
pressure coincide. The corresponding physical insight is that the
fluid may be divided into blocks that are in mechanical equilibrium
with each other and with any confining walls, so bulk and wall-based
pressure definitions must agree.


Purely thermodynamic concepts, like temperature, are well known to be
ill-defined in systems far from
equilibrium~\cite{Cugliandolo2011}. However, one could hope that
mechanical properties, like pressure, are less problematic. Here we
investigate this question for active fluids, in which energy
dissipation at the microscopic level drives the motion of each
particle to give strong non--equilibrium
effects~\cite{MarchettiRMP2013}. Assemblies of self-propelled
particles (SPPs) have been proposed as simplified models for systems
ranging from bacteria~\cite{Bergbook,CatesRRP2012} and active
colloidal
`surfers'~\cite{PalacciPRL2010,FilyPRL2012,ButtinoniPRL2013}, to
shaken grains~\cite{KudrolliPRL2008,NarayanSCI2007} and bird
flocks~\cite{BalleriniPNAS2008}.
We define the mechanical pressure $P$ of an active fluid as the mean
force per area exerted by its constituent particles on a confining
wall. This was studied numerically for a number of active systems,
showing some surprising effects for finite-size, strongly confined
fluids
\cite{MalloryPRE2014,FilySM2014,Fily2014,Niarxiv2014,Yang2014,TakatoriPRL2014,Takatoriarxiv2014}. Alternatively,
when describing the dynamics of such active fluids at larger scales,
some authors have introduced a bulk stress tensor and defined pressure
as its
trace~\cite{MarchettiRMP2013,Yang2014,TakatoriPRL2014,Takatoriarxiv2014},
leading to recent experimental measurements~\cite{GinotPRX2014}.
{ Since we are far from equilibrium, an equivalence between these
  different definitions, as seen numerically
  in~\cite{MalloryPRE2014,Yang2014,TakatoriPRL2014}, {requires
    explanation.}

  {In this article, we show analytically and numerically that the
    pressure $P$ exerted on a wall by generic active fluids
    {{\em directly depends}} on the microscopic interactions between
    the fluid and the wall. {Unless these interactions, as well
        as the interactions between the fluid particles}, obey strict
      and exceptional criteria, there is no equation of state
    relating the mechanical pressure to bulk properties of the
    fluid. Therefore, all connections to thermodynamics and to {the}
    bulk stress tensor are lost. Nevertheless, we provide analytical
    formulas to compute the wall-dependent pressure for some of the
    most studied classes of active systems. {Exceptional models for
      which an equation of state is recovered include the strictly
      spherical SPPs considered
      in~\cite{MalloryPRE2014,Yang2014,TakatoriPRL2014}.  Below we
      find that such simplified models are structurally unstable:
      small orientation-dependent interactions (whether wall-particle
      or particle-particle) immediately destroy the equation of
      state. Such interactions are present in every experimental
      system we know of.}

    {A clear distinction exists between the present work and that
      of Ref 20. The latter includes an explicit proof that pressure
      is, after all, well defined within a narrow class of models:
      spherical SPPs with torque-free wall interactions and
      torque-free pairwise interparticle forces. Because this class
      has been a major focus of theory and simulation studies, that
      finding is important, creating in those cases a direct link
      between pressure and correlation functions that can be exploited
      in future theoretical advances.  However, in general terms it is
      even more important to know that an equation of state for the
      pressure is the exception, rather than the rule, in active
      matter systems. This we establish here.}

To appreciate the remarkable consequences of the generic {absence of an}
equation of state, consider the quasi-static compression of an active
fluid by a piston.} Since the mechanical pressure depends on the
piston, compressing with a very soft wall---into which particles bump
gently---or with a very hard one requires different forces and hence
different amounts of work to reach the same final density. 
{This is not the only way our thermodynamic intuition can fail for
  active systems. We will show both that pressure can be anisotropic,
  and that active particles admit flux-free steady-states in which the
  pressure is inhomogeneous. Finally,} in the models we consider
(which best describe, e.g., crawling bacteria~\cite{Bergbook} or
colloidal surfers or rollers near a supporting
surface~\cite{PalacciSCI2014,BricardNAT2013}) there are situations in
which the confinement forces at the edges of a sample do not sum to
zero. We show how this unbalanced force is compensated by momentum
transfer to the support.} {The issue of whether an equation of
state exists in so-called ``wet'' active
matter~\cite{MarchettiRMP2013}---in which full momentum conservation
applies throughout the interior of the system---remains open.}

\section*{Non--interacting particles}
We consider a standard class of models for SPPs in which the independent Brownian motion of
each particle (diffusivity $D_t$) is supplemented by self-propulsion at speed $v$
in direction ${\bf u}$,
\begin{equation}\label{eqn:apdyn}
  \frac{d{\bf r}}{dt}= v {\bf u} + \sqrt{2 D_t}~{\boldsymbol \eta}(t)\;,
\end{equation}
with ${\boldsymbol \eta}(t)$ a Gaussian white noise of unit variance. The
reorientation of the direction of motion ${\bf u}$ then occurs with a
system-specific mechanism: active Brownian particles (ABPs) undergo
rotational diffusion, while run-and-tumble particles (RTPs) randomly undergo complete reorientations (`tumbles') at a certain rate. These well-established models have been
used~\cite{FilyPRL2012,StenhammarPRL2013,RednerPRL2013,SchnitzerPRE1993,TailleurPRL2008,CatesRRP2012,CatesPNAS} to describe respectively active
colloids~\cite{PalacciPRL2010,ButtinoniPRL2013,BricardNAT2013,PalacciSCI2014}, or
bacterial motion~\cite{Bergbook,CatesPNAS} and cell migration~\cite{RTPcells}. Such models neglect any coupling to a
momentum conserving solvent, and are thus best suited to describe
particles whose locomotion exploits the presence of a gel matrix or supporting surface as a momentum sink. This is true of many active systems, such as crawling
cells~\cite{SepulvedaPCB2013}, vibrated disks or grains~\cite{DeseignePRL2010,NarayanSCI2007,KudrolliPRL2008}, and colloidal
rollers~\cite{BricardNAT2013} or sliders~\cite{PalacciSCI2014}.

We address a system of SPPs with spatial coordinates ${\bf r} = (x,y)$
in 2D; we assume periodic boundary conditions, and hence translational
invariance, in the $\hat y$ direction. The system is confined along
$\hat x$ by two walls at specified positions, which exert forces
$-\nabla V(x)$ on particles at $x$; these forces have finite range and
thus vanish in the bulk of the system. The propulsion direction of a
particle is ${\bf u} = (\cos \theta,\sin\theta)$ with $\theta=0$ along
the $\hat x$ direction.  {In the absence of interactions between the
particles, the master equation for the probability $\cP({\bf
  r},\theta,t)$ of finding a particle at position ${\bf r}$ at time
$t$ pointing along the $\theta$ direction reads}
\begin{equation}
  \partial_t \cP = - \nabla \cdot [({\bf v} - \mu_t \nabla V(x)) \cP-D_t \nabla \cP] -\partial_\theta[\mu_r \Gamma(x,\theta) \cP- D_r \partial_{\theta}\cP]  -\alpha \cP +  \frac{\alpha}{2\pi}\int \cP\, d\theta' \; .
  \label{eq:nonint}
\end{equation}
Here $\mu_t$ and $D_t$ are the translational mobility and diffusivity;
likewise $\mu_r$ and $D_r$ for rotations. The propulsive velocity is
${\bf v}=v{\bf u}(\theta)$, and $\alpha$ is the tumble rate. ABPs
correspond to $\alpha=0$ and RTPs to $D_r=0$. Here we allow all
intermediate combinations, to test the generality of our results.  In
addition to the external force $-\nabla V(x)$, we include an external
torque $\Gamma(x,\theta)$, which may, for example, describe the
well--documented alignment of bacteria along
walls~\cite{ElgetiEPL2009}. Generically, just as in passive fluids, a
wall--torque will arise whenever the particles are not spherical
{and its absence is thus {strictly} exceptional. Obviously, the
  asphericity of (say) water molecules does not violate the
  thermodynamical precepts of pressure; remarkably, we show below
  that, for active particles, it does so.}

Since our setup is invariant along the $\hat{y}$ direction, the
mechanical pressure can be computed directly from the force exerted by
the system on a wall (which we place at $x = x_w\gg 0$), as
\begin{equation}
  \label{eq:mechanical_pressure}
  P=\int_{0}^\infty \rho(x) \partial_x V(x)\, dx \; .
\end{equation}
Here an origin $x=0$ is taken in the bulk, and $\rho(x)=\int_0^{2\pi}
\cP(x,\theta) d\theta$ is the steady-state density of particles at
$x$. As stated previously, for a passive equilibrium system ($v=0$)
with the same geometry, the mechanical
definition~\eqref{eq:mechanical_pressure} of pressure is equivalent to
the thermodynamic definition, as proved for completeness in the
Supplementary Information (SI).  Note that
Eq.~\eqref{eq:mechanical_pressure} still applies in the presence of
other particles, such as solvent molecules, so long as those particles
do not themselves exert any direct force on the wall (which is thus
semi-permeable). Under such conditions $P$ is, by definition, an {\em
  osmotic} pressure; the results below will still apply to it,
whenever Eq.(2) remains valid.

{As described in the SI, the pressure can be computed analytically
  from Eq.~\eqref{eq:nonint} as:}
\begin{equation}
  \label{eq:mechanical_pressure4}
  P = \left[\frac{v^2}{2\mu_t(D_r+\alpha)}+\frac{D_t}{\mu_t}\right] \rho_0 - \frac{v\mu_r}{\mu_t(D_r+\alpha)}\int_{0}^\infty dx \int_0^{2\pi}   \Gamma(x,\theta) \sin\theta \,\cP(x,\theta)\,
  d \theta\; .
\end{equation}
This is a central result, and exact for all systems obeying
Eq.~(\ref{eq:nonint}). Clearly, $\Gamma(x,\theta)$ in general depends
on the wall-particle interactions, as does $\cP(x,\theta)$ which is
sensitive to both $\Gamma(x,\theta)$ and $V(x)$. Thus the mechanical
pressure $P$ obeying Eq.~(\ref{eq:mechanical_pressure4}) is likewise
sensitive to these details: it follows that {\em no equation of state
  exists} for active particle systems in the general case.

\begin{figure}
  \begin{center}
    \includegraphics[width=.8\textwidth]{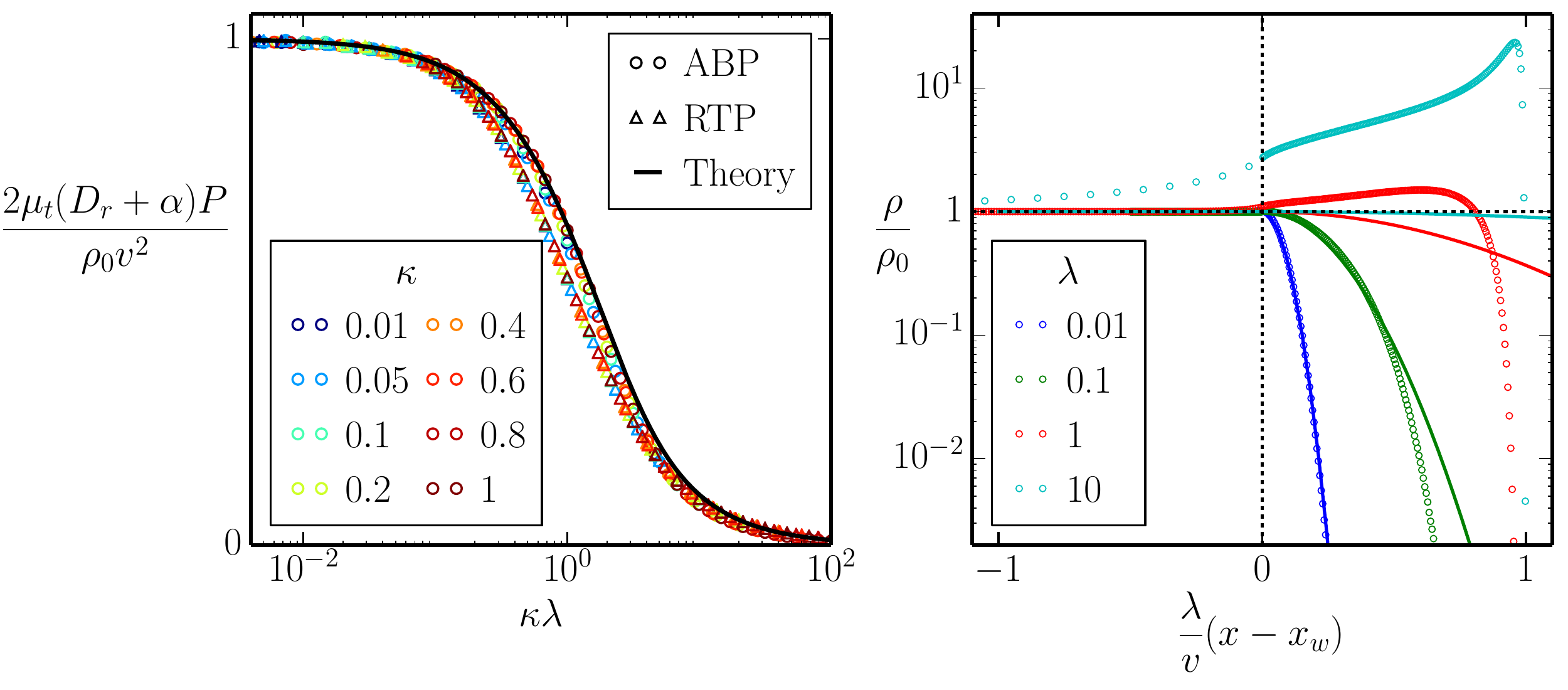}
    \caption{{\bf Non--interacting self-propelled ellipses.} {\bf
        Left:} Normalized pressure as the particle anisotropy $\kappa$
      and the wall stiffness $\lambda$ are varied for ABPs and
      RTPs. The theoretical prediction for ABPs correspond to
      Eq.~\eqref{eq:pressureABP} {\bf Right}: Density profiles for
      spherical particles for four different wall stiffness all
      yielding a pressure equal to $\rho k T_\text{eff}$. The full
      lines are Boltzmann distributions at $kT_\text{eff}$, showing
      that the pressure is given by the effective temperature far
      outside the Boltzmann regime $\lambda\ll D_r$. $v=D_r=1$ and
      $D_t=0$, with box size $L_x\times L_y=10\times 1$.}
    \label{fig:ABP-RTP_torque}
  \end{center}
\end{figure}

{To illustrate this effect and show
  that~\eqref{eq:mechanical_pressure4} can indeed be used to compute
  the pressure}, we study a model of ABPs with elliptical shape (see
SI for details). We choose a harmonic confining potential,
$V(x)=\frac{\lambda}{2}(x-x_w)^2 $ for $x>x_w$, with $V = 0$
otherwise, accompanied by a torque $\Gamma = \lambda\kappa\sin
2\theta$ (again, for $x>x_w$ and zero otherwise). With
$\kappa=(a^2-b^2)/8$, this is the torque felt by an elliptical
particle of axial dimensions $a,b$ and unit area $\pi a b$, subject to
the linear force field $-\nabla V(x)$ distributed across its body.
Assuming the steady-state distribution $\cP(x,\theta)$ to relax to its
bulk value outside the range of the wall potential,
$\cP(x_w,\theta)=\rho_0/2\pi$, the pressure in such an ABP fluid (for
$D_t=0$) is given by
\begin{equation} P= \frac{\rho_0 v^2}{2\lambda \mu_t\mu_r
    \kappa} \left[1-\exp\left(-\frac{\lambda\mu_r \kappa}{D_r}\right)
  \right]\; .
\label{eq:pressureABP}
\end{equation}
For $\kappa>0$ the torque reduces the pressure by orienting the ABPs
parallel to the wall.  Equation~\eqref{eq:pressureABP} shows
explicitly how walls with different spring constants $\lambda$
experience different pressures, in sharp contrast with thermodynamics.
We checked this prediction by direct numerical simulations of ABPs and
found good agreement (see Fig.~\ref{fig:ABP-RTP_torque}). We also
found similar behavior numerically for (likewise elliptical) RTPs,
confirming that the failure of thermodynamics is generic.

For passive particles in thermal equilibrium, $v=0$ and
Eq.~(\ref{eq:mechanical_pressure4}) reduces to the ideal gas law, $P =
\rho_0k_BT$, upon use of the Einstein relation ($D_t/\mu_t =
k_BT$). Another case where an equation of state is recovered is for
torque--free (e.g., spherical) particles, with $\Gamma = 0$.  In that
case Eq.~(\ref{eq:mechanical_pressure4}) reduces to the same ideal gas
law but with an effective temperature
\begin{equation}
  \label{eq:effective_temp}
  \frac{P}{\rho_0} = k_BT_\text{eff}=\frac{v^2}{2\mu_t(D_r+\alpha)}+\frac{D_t}{\mu_t}\;.
\end{equation}
This explains why previous numerical studies of torque--free,
non--interacting active particle fluids gave consistent pressure
measurements between impenetrable~\cite{FilySM2014,Fily2014} or
harmonically soft walls~\cite{MalloryPRE2014}. Related expressions for
the pressure of such fluids were found by computing the mean kinetic
energy~\cite{MalloryPRE2014}, or the stress
tensor~\cite{Yang2014,TakatoriPRL2014,Takatoriarxiv2014}, possibly
encouraging a belief that all reasonable definitions of pressure in
active systems are equivalent. However,
Eq.~\eqref{eq:mechanical_pressure4} shows that these approaches cannot
yield consistent results beyond the simplest, torque--free case.

The ``effective gas law" of Eq.~(\ref{eq:effective_temp}) for the
torque--free case is itself remarkable.  For ABPs or RTPs in an
external potential $V(x)$, the effective temperature concept predicts
a steady-state density $\rho(x)\propto \exp[-V(x)/k_BT_\text{eff}]$
that is accurate {\em only} for weak force
fields~\cite{TailleurEPL2009,ABPvsRTP}. Yet Eq.~(\ref{eq:effective_temp}) holds
even with hard-core walls for which the opposite applies and the
steady-state density profile is far from a Boltzmann distribution (see
the simulation results of Fig.~\ref{fig:ABP-RTP_torque} and the
analytical results for one--dimensional RTPs in SI). {In fact the
  result stems directly from the exact computation of $\int_0^\infty
  \rho(x) \partial_x V(x) \,d x$, which can be done at the level of the
  master equation and leads to Eq.~(\ref {eq:mechanical_pressure4}),
  so that no broader validity of the $T_\text{eff}$ concept is
  required, or implied.}

{
\section*{Interacting active particles}

Equation~\eqref{eq:mechanical_pressure4} gives the pressure of
non--interacting active particles and we now address the extent to
which our conclusions apply to interacting SPPs. Clearly, interactions
will not restore the existence of an equation of state in the presence
of wall torques and we thus focus on ``torque-free'' walls.

Interparticle alignment is probably the most studied interaction in
active matter~\cite{MarchettiRMP2013}. To measure its impact on pressure, we consider $N$
ABPs whose positions $\r_i$ and {orientations} $\theta_i$ evolve
according to~\eqref{eqn:apdyn} and
\begin{equation}
  \frac{d\theta_i}{dt}=\mu_r\sum_{j=1}^N F(\theta_j-\theta_i,\r_i,\r_j)+\sqrt{2 D_r} \xi_i(t)
\end{equation}
where $F$ is an aligning {torque} between the particles. As shown in SI, the
pressure can be computed analytically to give
\begin{equation}
  \label{eq:pressure-aligning}
  P = \left[\frac{v^2}{2\mu_tD_r}+\frac{D_t}{\mu_t}\right] \rho_0 - \frac{v\mu_r}{\mu_tD_r}\int_{0}^\infty  \!\!\!\!\! dx\int_{-\infty}^\infty  \!\!\!\!\! dy \int_0^{2\pi} \!\!\!\!\! d \theta\int  \! d\r' \int_0^{2\pi} \!\!\!\!\! d \theta' 
  F(\theta'-\theta,\r,\r') \sin\theta \langle \cP(\r,\theta)\cP(\r',\theta')\rangle 
\end{equation}
where the integral over $\r'$ is over the whole space. Since the
distribution $\cP(\r,\theta)=\sum_{i=1}^N
\delta(\r-\r_i)\delta(\theta-\theta_i)$ depends (for $x>0$) on the
wall potential, so does the pressure. Therefore, even in the absence
of wall torques, {alignment interactions between particles destroy any} equation of
state. Figure~\ref{fig:pressure_interacting} shows the result of ABP
simulations with a particular choice of interparticle torque $F$: The
measured pressure indeed depends on the wall potential and agrees with
equation~\eqref{eq:pressure-aligning}.

\begin{figure}
  \begin{center}
    \includegraphics[width=.32\textwidth]{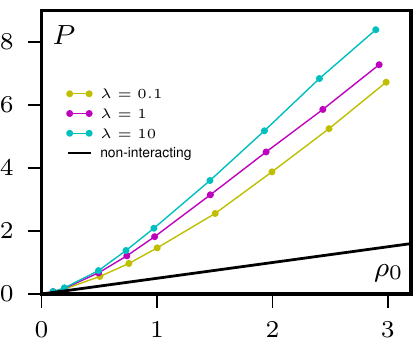}
    \includegraphics[width=.32\textwidth]{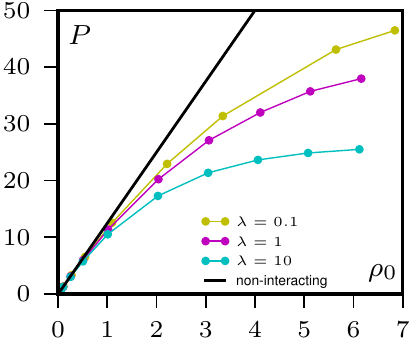}
    \includegraphics[width=.32\textwidth]{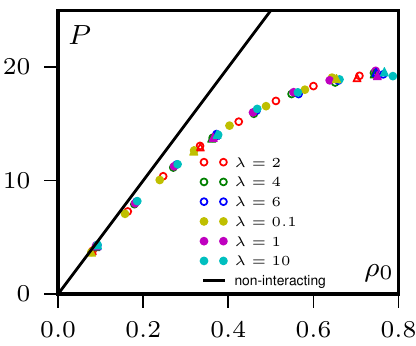}
    \caption{{{\bf Interacting self-propelled spheres.} Pressure
        versus density $P(\rho_0)$ for interacting particles
        ($L_x\times L_y=200\times 50$). {\bf Left:} Aligning ABPs. The
        torque exerted by particle $j$ on particle $i$ is
        $F(\theta_j-\theta_i,\r_i,\r_j)=\frac{\gamma}{\mathcal{N}(\r_i)}\sin(\theta_j-\theta_i)$
        if $|\r_j-\r_i|<R$ and $0$ otherwise, where
        $\mathcal{N}(\r_i)$ is the number of particles interacting
        with particle $i$. $v=1$, $D_r=1$, $D_t=0$, $R=1$ and
        $\gamma=2$.  {\bf Center}: ``Quorum sensing'' interactions
        $v(\bar\rho)=v_0(1-{\bar\rho}/{\rho_m})+v_1$ with $v_0=10$,
        $v_1=1$, $\rho_m=5$, $D_r=D_t=1$. {\bf Right} The pressure of
        particles interacting with repulsive WCA potentials is
        independant of the wall potential. Triangles and circles
        represent RTPs and ABPs with $v=10$, $D_r=1$, $\alpha=1$ and
        $D_t=0$. Open and full symbols correspond to linear and
        harmonic wall potentials. (See SI for numerical details.)}}
    \label{fig:pressure_interacting}
  \end{center}
\end{figure}

In active matter, more general interactions than pairwise torques
often have to be considered.  For example, in bacteria with ``quorum
sensing" (a form of chemical communication), particles at position
${\bf r}$ can adapt their dynamics in response to changes in the local
coarse--grained particle density $\bar\rho({\bf
  r})$~\cite{LiuSCI2011}. Also shown in
Fig.~\ref{fig:pressure_interacting} are simulations for the case
$v(\bar\rho)=v_0(1-{\bar\rho}/{\rho_m})+v_1$, reflecting a pairwise
speed {reduction} {(see SI for
details)}. {This is an example where} even completely torque-free
particles have no equation of state. Again, we show in SI how an
explicit formula for the pressure can be computed from first
principles.

The case of torque--free ABPs with short range repulsive
interactions~\cite{RednerPRL2013,StenhammarPRL2013,LowenEPL2013,GompperEPL2013}
was recently considered in~\cite{Yang2014,TakatoriPRL2014}. {The}
mechanical force exerted on a wall was found to coincide with a
pressure {computed from the bulk} stress tensor, {suggesting that in
  this case an equation of state does} exist. To check this, we choose
a Weeks--Chandler--Andersen (WCA) potential: $U(r)=4\left[
  \left(\frac{\sigma}{r}\right)^{12}
  -\left(\frac{\sigma}{r}\right)^{6}\right]+1$ if $r<2^{1/6}\sigma$
and $U=0$ otherwise, where $r$ is the inter-particle distance and
$\sigma$ the particle diameter. Using simulations we determined $P$ as
a function of bulk density $\rho_0$ for various harmonic and linear
wall potentials. As shown in Fig.~\ref{fig:pressure_interacting}, all
our data collapses onto a wall-independent {equation of state
  $P(\rho_0)$. {An analytical expression for $P(\rho)$ in this rather
  exceptional case is derived and studied in~\cite{pressureABP} in the
  context of phase equilibria.}}

The cases explored above {show that there is generically no
equation of state in an active fluid, one exception being when wall-particle and particle-particle torques are both negligible.}
Given {this outcome}}, a simple test for
the presence or absence of an equation of state, in simulations or
experiments, would be welcome. If the pressure is set by bulk
properties of the fluid, when an asymmetrically interacting partition
is used to separate the system in two parts, no force acts upon the
partition and it does not move. Conversely, if the partition does
move, there is no equation of state. To check this, we simulated a
large box of homogeneous active fluid, introduced at its centre a
mobile wall with asymmetric potentials on its two sides, and let the
system reach steady state. In the cases shown above to have an
equation of state, the wall remains at the center of the box so that
the densities on its two sides stay equal. In the other cases,
however, the partition moves to equalize the two wall--dependent
pressures, {resulting in a flux-free steady-state with unequal
densities in the two chambers (Fig.~\ref{piston})}.

\begin{figure}
  \begin{center}
    \includegraphics[width=.49\textwidth]{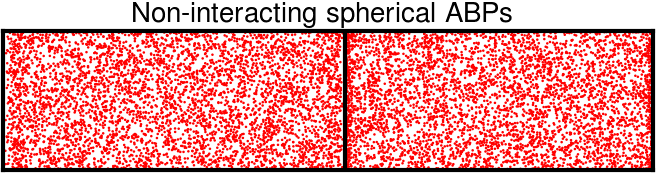}\includegraphics[width=.49\textwidth]{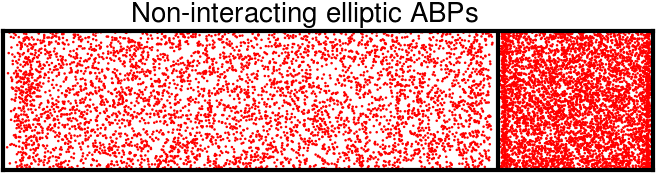}

    \includegraphics[width=.49\textwidth]{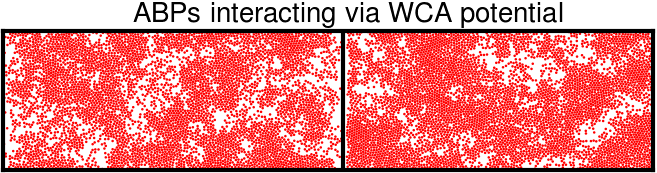}\includegraphics[width=.49\textwidth]{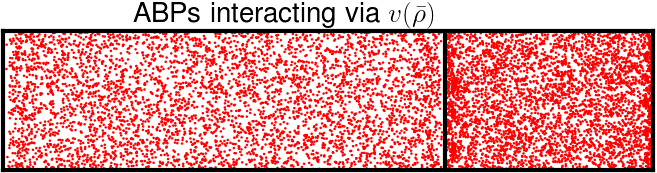}
      \caption{{\bf Simple test for the existence of an equation of state.} Four
        snapshots of the steady-state of $10\,000$ ABPs in a $200\times
        50$ cavity split in two by a mobile asymmetric harmonic wall
        ($\lambda=1$ on the left and $\lambda=4$ on the right, $v=10$,
        $D_r=1$, $D_t=0$) for: non--interacting spherical ABPs (top left),
        non--interacting elliptic ABPs with $\mu_r=\kappa=1$ (top right),
        ABPs interacting via the WCA potential (bottom left) and via
        $v(\bar \rho)$ (bottom right) with $v_0=10$, $v_1=1$,
        $\rho_m=4.8$. A spontaneous compression of the right half of the
        system is the signature of the lack of equation of state.}
  \label{piston}
\end{center}
\end{figure}

\section*{Anomalous attributes of the pressure}
{A defining property of equilibrium {\em fluids} is that they cannot
statically support an anisotropic stress.} Put differently, the normal
force per unit area on any part of the boundary is independent of its
orientation. This applies even to oriented fluids (without positional
order), such as nematic liquid crystals~\cite{ChaikinLubensky}, but
breaks down for active nematics~\cite{MarchettiRMP2013}.

\begin{figure}
  \begin{center}
    \includegraphics[width=0.5\textwidth]{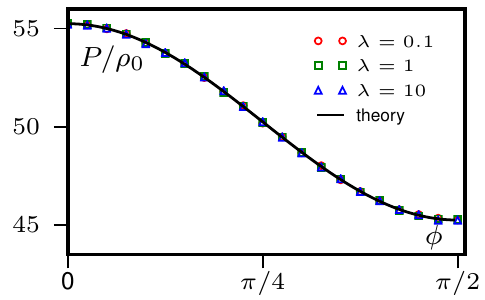}
  \caption{{\bf Anisotropic pressure.} RTPs with anisotropic speed $v(\theta)=v_0
    +v_1 \cos(2\theta)$, with $v_0=10$, $v_1=1$, $D_t=0$. The pressure
    depends on the angle $\phi$ between the wall and the axis $\hat y$
    but not on the stiffness of the potential.}
  \label{anisotropic}
\end{center}
\end{figure}

We next show that it can also break down for active fluids with
isotropic particle orientations, as long as the propulsion speed is
anisotropic, i.e. $v=v(\theta)$. This could stem from an anisotropic
mobility $\mu_t(\theta)$ as might arise for
cells crawling on a corrugated surface. We suppose $v(\theta) =
v(\theta+\pi)$ so that oppositely oriented particles have the same
speed; Eq.~(\ref{eq:nonint}) then shows that the bulk steady state
particle distribution ${\cal P}({\bf r},\theta)$ remains isotropic. In
addition, as shown in SI, the pressure $P(\phi)$ acting on a wall
whose normal is at angle $\phi$ to the $\hat x$ axis remains
independent of the wall interactions, but is $\phi$-dependent; for
RTPs ($D_r=0$) it obeys
\begin{equation}
  \label{eq:non-isotropic}
  P(\phi)= \frac{\rho_0 D_t}{\mu_t}+ \frac{\rho_0}{2\pi \mu_t  \alpha} \int_0^{2\pi} v^2(\theta) \cos^2 (\theta-\phi)\, d\theta\ \;.
\end{equation}
To verify that the pressure is indeed anisotropic we performed numerical simulations for
$v(\theta)= v_0+v_1\cos(2\theta)$ which show perfect agreement with
Eq.~\eqref{eq:non-isotropic} (see Fig.~\ref{anisotropic}).

For passive fluids without external forces, mechanical equilibrium
requires that the pressure is not only isotropic, but also uniform. This follows from the Navier--Stokes equation
for momentum transport~\cite{ChaikinLubensky}, but also holds in (say) Brownian dynamics simulations which do not conserve momentum~\cite{Allen}.

\begin{figure}
\raisebox{.35cm}{\includegraphics[width=.4\textwidth,totalheight=2.5cm]{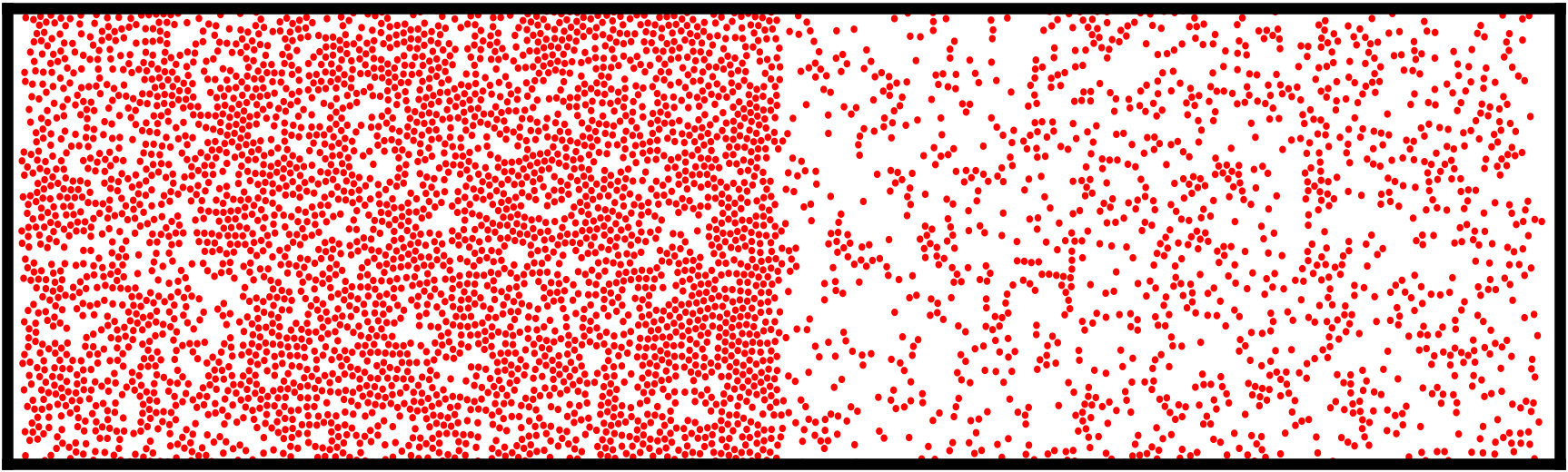}} \includegraphics[width=.3\textwidth]{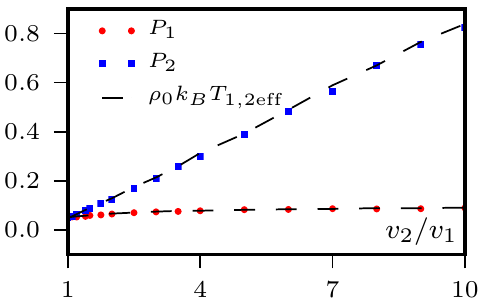}\includegraphics[width=.3\textwidth]{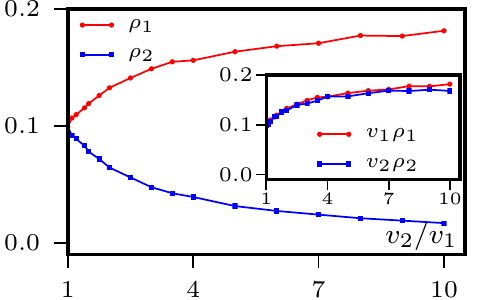}
\caption{{\bf Inhomogeneous pressure}. Spherical ABPs interacting
    with WCA potential, with speeds $v_1$ for $x<0$ and $v_2$ for
    $x>0$. {\bf Left:} Snapshot of the cavity in steady state
    ($v_1=1$, $v_2=5$). {\bf Middle:} Pressures $P_1$ and $P_2$ as
    $v_2/v_1$ is varied. {\bf Right:} As $v_2/v_1$ varies, the
    densities evolve to equalize $\rho v$ rather than $P\sim\rho
    v^2$. $D_r=1$, $D_t=0$, $\lambda=1$, $L_x\times L_y=200\times
    50$.}
\label{fig:v1v2}
\end{figure}

We now show that $P$ need not be uniform in active fluids, even when
an equation of state exists. Consider non--interacting spherical ABPs
in a closed container with different propulsion speeds in different
regions, say $v=v_1$ for $x<0$ and $v=v_2$ for $x>0$. This is a
realizable laboratory experiment in active colloids whose propulsion
is light-induced~\cite{PalacciSCI2014,ButtinoniPRL2013}. From
Eq.~\ref{eq:nonint}, the flux-free steady state has $\rho \propto 1/v$
throughout~\cite{SchnitzerPRE1993,TailleurPRL2008,CatesEPL2013}, so
that the pressures $P_{1,2}\propto \rho v^2$ are unequal.
Though different, the pressures in the two compartments are well
defined, uniform within each bulk, and independent of the
wall-particle interactions. They remain different when interparticle
interactions are added (see Fig~\ref{fig:v1v2}). Indeed, if for
$v_1\neq v_2$ equality of the ideal pressure is restored by setting
$D_{r} \propto v$, the effect of such interactions is to reinstate a
pressure imbalance. Nonuniformity of $P$ is thus fully generic for
nonuniform $v$.

The above example implies a remarkable result, that also holds for
systems with no equation of state enclosed by spatially heterogeneous
walls. In both cases {\em the net force acting across the system
  boundary is generically nonzero}. Were momentum conserved, this
would require the system as a whole to be accelerating. Recall however
that Eq.~(\ref{eq:nonint}) describes particles moving on, or through,
a medium that absorbs momentum and this net force is exactly cancelled
by the momentum exchange with the support. The latter vanishes on
average in the isotropic bulk, but is nonzero in a layer of finite
polarization ($m_1\neq 0$) close to each wall.

\section*{Discussion}

{Our work shows that in active fluids the concept of pressure defies
  many suppositions based upon concepts from thermal equilibrium. The
  generic absence of an equation of state is the most striking
  instance of this.  {Despite its absence, we have shown how to
    compute the mechanical pressure for a large class of active
    particle systems. Clearly, the concept of pressure is even more
    powerful in the exceptional cases where an equation of state does
    exist. This excludes any chemically-mediated variation in
    propulsion speed, and also requires wall--particle and
    interparticle torques both to be negligible.  {Because it can
      easily be achieved on a computer, though not in a laboratory,
      the torque-free case of spherical active Brownian particles
      without bulk momentum conservation has played a pivotal role in
      recent theoretical studies of active
      matter~\cite{MalloryPRE2014,Yang2014,TakatoriPRL2014}. The
      proof~\cite{pressureABP} that an equation of state does exist
      for this system is all the more remarkable because, as we have
      seen, such an outcome is the exception and not the rule.}

It is interesting to inquire how our results would change for systems
with full momentum conservation in the bulk. As mentioned previously,
if Eq.~\eqref{eq:nonint} still applies, our exact results for $P$
remain valid so long as this is taken as an {\em osmotic} pressure.
For dilute systems Eq.~\eqref{eq:nonint} should indeed hold in bulk,
even though particles now propel by exerting force multipoles on the
surrounding solvent. (Since the walls of the system are semipermeable,
the solvent can carry momentum across them, and effectively becomes a
momentum sink for the active particles.)  However, even for spherical
swimmers, hydrodynamic interactions can now cause torques, {both
  between the particles and} near the wall~\cite{BerkePRL2008}, making
an equation of state less likely. Its absence would then manifest as a
nonzero net force on a semipermeable partition between two identical
samples of (say) a swimming bacterial fluid. We predict this outcome
whenever the two faces of the partition have different interactions
with the swimming particles.

\bibliographystyle{Science}

\begin{acknowledgements}
 We thank K. Keren, M.  Kolodrubetz, C. Marchetti,
  A. Polkovnikov, J.  Stenhammar, R. Wittkowski and Xingbo Yang for
  discussions.  This work was funded in part by EPSRC EP/J007404. MEC
  holds a Royal Society Research Professorship. YK was supported by
  the I-CORE Program of the Planning and Budgeting Committee and the
  Israel Science Foundation. MK is supported by NSF grant
  No. DMR-12-06323. AB and YF acknowledge support from NSF grant
  DMR-1149266 and the Brandeis Center for Bioinspired Soft Materials,
  an NSF MRSEC, DMR-1420382. Their computational resources were
  provided by the NSF through XSEDE computing resources and the
  Brandeis HPCC. YK, AS and JT thank the Galileo Galilei Institute for
  Theoretical Physics for hospitality. AB, MEC, YF, AS, MK and JT
  thank the KITP at the University of California, Santa Barbara, where
  they were supported through National Science Foundation Grant NSF
  PHY11-25925.
\bibliography{scibib}
\end{acknowledgements}

\newpage

\begin{center}
  \LARGE Supplementary information
\end{center}
\section{Details of Numerical Simulations}

{\bf Time-stepping}: Simulations were run using Euler
time-discretization schemes over total times $T=10^4$ or larger (up to
$T=10^9$).

{\bf Non-interacting particles}: At each time step $dt$, particles
update their direction of motion $\theta_i$, then their position ${\bf
  r_i}$. For ABPs, $\dot\theta_i=\sqrt{2D_r}\xi(t)$ where $\xi(t)$ is
a Gaussian white noise of unit variance. For RTPs, the time $\Delta t$
before the next tumble is chosen using an exponential distribution
$P(\Delta t)=\lambda e^{-\lambda \Delta t}$. When this time is
reached, a new direction is chosen uniformly in $[0,2\pi[$ and the
next tumble time is drawn from the same distribution. This neglects
the possibility to have two tumbles during $dt$. Both types of
particles then move according to the Langevin equation $\dot {\bf
  r_i}=v {\bf e_{\theta_i}}-\nabla V +\sqrt{2 D_t}\eta(t)$ where
$\eta(t)$ is a Gaussian white noise of unit variance. 

{\bf Hard-core repulsion}: To model hard-core repulsion we use a WCA
potential $V(r)=4\big[ \left(\frac{\sigma}{r}\right)^{12}
-\left(\frac{\sigma}{r}\right)^{6}\big]+1$ if $r<2^{1/6}\sigma$ and 0
otherwise. The unit of length is chosen such that the interaction
radius $2^{1/6}\sigma=1$. Because of the stiff repulsion, one needs to
use much smaller time steps ($dt=5.10^{-5}$ for the speeds considered
in the paper).

{
{\bf Aligning particles}: Particles exert torques on each other to
align their directions of motion $\theta_i$. The torque exerted by
{particle} $j$ on {particle} $i$ reads $F(\theta_j-\theta_i,\vec
r_j-\vec r_i)=\frac{\gamma}{\mathcal{N}(\vec
  r_i)}\sin(\theta_j-\theta_i)$ if $|\vec r_i-\vec r_j|<R$ and $0$
otherwise, where $\mathcal{N}(\vec r_i)$ is the number of particles
interacting with particle $i$. The interaction radius $R$ is chosen as
unit of length.  For the parameters used in simulations $v=1$,
$\gamma=2$, with a time-step $dt=10^{-2}$.}

{\bf Quorum sensing $v(\bar\rho)$}: The velocities of the particles
depend on the local density $\bar\rho$. The unit of length is fixed
such that the radius of interaction is 1. To compute the local
density, we use the Schwartz bell curve
$K(r)=\frac{1}{Z}\exp(-\frac{1}{1-r^2})$ for $r<1$ and $0$ otherwise,
where $Z$ is a normalization constant. The average density around
particle $i$ is then given by $\bar\rho_i=\sum_j K(|{\bf r_i-r_j}|)$
and the velocity of particle $i$ is
$v(\bar\rho_i)=v_0(1-{\bar\rho_i}/{\rho_m})+v_1$.  We used
$dt=5.10^{-3}$.

{\bf Asymmetric wall experiment}: The simulation box is separated in
two parts by an asymmetric wall which has a different stiffness
$\lambda_1$ and $\lambda_2$ on both sides. At each time step, the
total force $\mathcal F$ exerted on the wall by the particles is
computed and the wall position is updated according to $\dot x_{\rm
  wall}=\mu_{\rm wall} \mathcal F$, where $\mu_{\rm wall}=2.\,10^{-4}
\ll \mu_t$ is the wall mobility.

{\bf SI movie 1}: Asymmetric wall experiment with non-interacting ABP
particles. The particles are spherical (no torque) for $t<1000$ and
$t>3000$ and ellipses with $\kappa=1$ for $1000<t<3000$. Wall
potentials are harmonic and other parameters are $v=10$, $D_r=1$,
$\lambda=10$ (external box) and for the asymmetric mobile wall
$\lambda=1$ on the left and $\lambda=4$ on the right.

\section{Equilibrium Pressure}
Here, for completeness, we show that in equilibrium 1) the
thermodynamic pressure equals the mechanical pressure given by Eq.~(3)
of the main text, and 2) that it is independent from the wall
potential. For simplicity we consider a system of interacting
point-like particles in one-dimension where the pressure is a force
and we work in the canonical ensemble. The extension to other cases is
trivial.

The thermodynamic  pressure is defined as 
\begin{equation}
P=-\left.\frac{\partial F}{\partial L}\right|_N\;,
\end{equation}
where $L$ is the system length, $F$ is the free energy, and the number
of particles $N$ is kept constant.  Note that since $F$ is extensive,
any contribution from the potential of the wall is finite and will
therefore not influence the pressure. Next, the free energy is given
by
\begin{equation}
F=-\frac{1}{\beta} \ln {\cal Z}\;,
\end{equation}
where
\begin{equation}
{\cal Z} = \sum_n e^{-\beta [( {\cal H} + \sum_i V(x_i-L)]}\;,
\end{equation}
is the partition function, $\beta=1/T$ with $T$ the temperature, and
the sum runs over all micro-states. The origin of the wall is chosen
at $x=L$, as opposed to $x=x_w$ in the main text. The energy function
of the system is given by ${\cal H}+\sum_iV(x_i-L)$, where $V(x_i-L)$
is the wall potential, $x_i$ is the position of particle $i$, and ${\cal
  H}$ contains all the other interactions in the system. Using the
definition of $P$ we have
\begin{equation}
{ P}=-\frac{1}{{\cal Z}} \sum_n \sum_i \partial_LV(x_i-L) e^{-\beta ( {\cal H} +\sum_i V(x_i-L))} = -\left\langle \int dx \rho(x)  \partial_L V(x-L) \right\rangle\;,
\end{equation}
where the angular brackets denote a thermal average, and
$\rho(x)=\sum_i \delta(x-x_i)$ is the number density. Exchanging
$\partial_L$ for $-\partial_x$, we obtain the expression from the main
text
\begin{equation}
P=\left\langle \int dx \rho(x)  \partial_x V(x-L) \right\rangle \;.
\end{equation}

\section{Derivation of the pressure for non-interacting SPPs}
\label{sec:derivationP}
To compute the mechanical pressure $P$ for SPPs, we first define
  $m_n(x)=\int_0^{2\pi} \cos(n \theta)\cP(x,\theta) d\theta$.
  Taking moments of the master equation, Eq.~(2) in the main text, we
  find that in steady state
\begin{align}
  \label{eq:ss-rho}
  0&=-\partial_x (v m_1-\mu_t \rho \partial_xV-D_t\partial_x\rho) \;,\\
  \label{eq:ss-m}
  (D_r+\alpha)m_1&=-\partial_x \Big(v\frac{\rho+m_2}{2} -\mu_t m_1 \partial_xV-D_t \partial_x m_1\Big)- \int_0^{2\pi}\sin\theta\, \mu_r\Gamma(x,\theta) \cP\,d\theta\;.
\end{align}
Equation~(\ref{eq:ss-rho}) is tantamount to setting $\partial_x J = 0$, where $J$ is a particle current that must vanish in any confined system; while Eq.~(\ref{eq:ss-m}) expresses a similar result for the first moment $m_1$.
Equation~(3) of the main text and Eq.~(\ref{eq:ss-rho}) together imply that 
\begin{equation}
  \label{eq:mechanical_pressure2}
  P=\int_{0}^\infty \frac{1}{\mu_t} \left[v m_1-D_t\partial_x\rho\right] \, dx \;.
\end{equation}

Next, from Eqs.~(\ref{eq:ss-rho},\ref{eq:ss-m}) we see that, apart
from the term involving the torque $\Gamma$, $m_1(x)$ is a total
derivative. We can trivially integrate this contribution to
Eq.~(\ref{eq:mechanical_pressure2}), noting that at $x=0$, isotropic
bulk conditions prevail so that $m_1=m_2=0$, and $\rho = \rho_0$
(say), while as $x\to\infty$, far beyond the confining wall,
$\rho=m_1=m_2=0$. Restoring the $\Gamma$ term we finally obtain
Eq.~(4) of the main text.

\section{Pressure for an ellipse in a harmonic potential}

In what follows we first compute the torque applied on an ellipse in a
harmonic potential. We then derive an approximate expression for the
pressure, Eq.~(5) of the main text, which is valid as long as the
density distribution $P(r,\theta)$ equals its bulk value as soon as
the wall potential vanishes (at $x=x_w$).

\subsection{Torque on an ellipse}

We consider an ellipse of uniform density and long and short axes of
lengths $a$ and $b$ respectively. We define two sets of axes: 1)
($\hat{x}$, $\hat{y}$) are the real space coordinates with the wall
parallel to the $y$ axis, and 2) ($\hat{x}_p$, $\hat{y}_p$) are the
coordinates associated with the ellipse so that $x_p$ is parallel to
its long axis. The angle between the two sets of coordinates is
$\theta$, which is also the direction of motion of the particle (see
Fig.~\ref{fig:ellipse}). For simplicity, we assume that the particle
is moving along its long principal axis.

\begin{figure}
\begin{center}
\begin{tikzpicture}
  \begin{scope}[rotate=18.66]
   \draw (0,0) ellipse (2 and 1);
   \draw[->,thick] (-2.5,0) -- (2.5,0) node (xaxis) [right] {$\hat{x}_p$};
   \draw[->,thick] (0,-1.5) -- (0,1.5) node (yaxis) [right] {$\hat{y}_p$};
   \draw[->,red,dashed] (-2.5,0.8) -- (2.5,-0.8) node (xpaxis) [right] {$\hat{x}$};
   \draw[->,red,dashed] (-0.48,-1.5) -- (0.48,1.5) node (ypaxis) [right] {$\hat{y}$};
   \draw[red,thick,<-] (1,0) arc (0:-18.66:1) node [below] {$\theta$};
 \end{scope}
\end{tikzpicture}
\caption{An illustration of the axes ($\hat{x}$, $\hat{x}$) and ($\hat{x}_p$, $\hat{x}_p$), and the angle $\theta$.}
\label{fig:ellipse}
\end{center}
\end{figure}
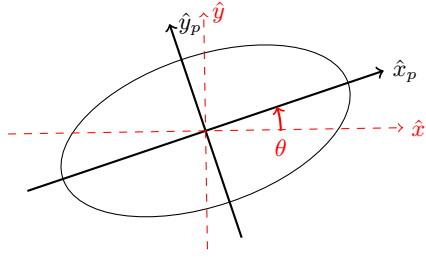

Since the wall is perpendicular to the $\hat{x}$ axis, the force acting
on an area element of the ellipse is given by
$F_w(x_0+x)=-\partial_{x} V(x_0+x)$, where $x_0$ is the position of the
center of mass of the ellipse and $x$ the relative coordinate of the
area element within the ellipse, both along the $\hat{x}$ direction.

The torque applied by the force at a point $\vec r$ is then given by
\begin{align}
  \label{eq:torque-point}
  \gamma &=\vec r \times F_w(x_0+x) \hat{x} \;,\\
  &=\colvec{x_p}{y_p} \times F_w(x_0+ x_p\cos\theta- y_p\sin \theta) \colvec{\cos\theta}{-\sin\theta}\;.
\end{align}

Next, we integrate over the ellipse, taking its mass density to be
uniform $\rho(x_p,y_p)=m/(\pi a b)$. Rescaling the axes as $x_p'=x_p/a$
and $y_p'=y_p/b$ to transform the ellipse into a unit circle, and
switching from $(x_p',y_p')$ to polar coordinates $(r,\varphi)$, yields
\begin{align}
  \label{eq:torque1}
  \Gamma&=\frac{m}{\pi a b}\int dx_p dy_p \gamma \\
  &=\frac{m}{\pi}\int dx_p' \int dy_p' F_w(x_0+ a x_p'\cos\theta- b y_p'\sin \theta) \colvec{ax_p'}{by_p'} \times  \colvec{\cos\theta}{-\sin\theta} \nonumber\\
  &=m\int_0^{2\pi}\frac{d\varphi}{\pi}\int_0^1dr r F_w(x_0+ a r \cos\varphi\cos\theta- b r\sin \varphi \sin \theta) \colvec{ar\cos \varphi}{br\sin \varphi} 
\times  \colvec{\cos\theta}{-\sin\theta}\;. \nonumber
\end{align}

For a harmonic wall potential $F_w(x)=-\lambda x$, the integral can be
computed, and we get
\begin{equation}
  \label{eq:torque_harmo}
  \Gamma=\frac{m\lambda}{8}(a^2-b^2)\sin(2\theta) \equiv \lambda \kappa \sin(2\theta)\;,
\end{equation}
which has the expected symmetries: it vanishes for a sphere ($a=b$),
and for particles moving along or perpendicular to the $x$-axis. Note
that the torque is constant, independent of the position of the
particle as long as the whole ellipse is within the range of the wall
potential. In the main text we assume that this is always the case,
which means that the ellipse is very small when compared to the
typical decay length of $\rho(x)$ due to $V$. In the simulations, we
thus simulated point-like ABPs with external torques $\Gamma=\pm
\lambda\kappa\sin 2\theta$ for left and right walls. For real systems,
the collision details would clearly be different, hence giving
different quantitative predictions for the pressure $P$, but the
qualitative results would be the same. We set $m=1$ for ease of
notation and define the asymetry coeficient $\kappa=(a^2-b^2)/8$ as in
the main text.

\subsection{Approximate expression for the pressure} 
We now turn to the derivation of the approximate expression Eq.~(5) in the
main text for the pressure. In particular we focus on the case of ABP
($\alpha=0$) ellipses confined by a harmonic wall potential and for
simplicity neglect the translational diffusion $D_t=0$. In that case
the contribution of the torque to the pressure reads
\begin{equation}
  C=\frac{ \bar \lambda  v}{\mu_t}\int_{0}^{+\infty}dx\int_0^{2\pi}d\theta  \sin(\theta)\sin(2\theta)\cP(x,\theta)\;,
\end{equation}
where we have used expression~\eqref{eq:torque_harmo} for $\Gamma$ and
defined $\bar\lambda=\mu_r\kappa\lambda/D_r$.

We will now expand the pressure $P$ as a power series in $\bar
\lambda$. If we make the approximation $P(x_w,\theta)=\rho_0/(2\pi)$,
so that the steady-state distribution relaxes to its bulk value as
soon as the system is outside the range of the wall potential, we can
resum the series to obtain Eq.~(5) of main text.

We first expand the probability distribution $\cP(x,\theta)$ in powers
of $\bar\lambda$
\begin{equation}
  \cP(x,\theta)=\sum_{k=0}^\infty \bar\lambda^k \cP_k(x,\theta)\;,
\end{equation}
so that the pressure is given by
\begin{align}
  \label{eq:pressureseries}
  P =\frac{v^2}{2\mu_t D_r}\rho_0-C =\frac{v^2}{2\mu_t D_r}\rho_0-\frac{v}{\mu_t}\sum_{k=0}^\infty C_k\bar\lambda^{k+1}\;,
\end{align}
where
\begin{equation}
  \label{eq:defC}
  C_k=\int_{x_w}^\infty dx\int_0^{2\pi}d\theta \sin\theta\sin(2\theta)\cP_k(x,\theta)\;.
\end{equation}

\subsubsection{Computation of the coefficients $C_k$}
$C_0$ is known since $\cP_0=\rho_0/2\pi$. Using the hypothesis
$\cP(x_w,\theta)=\rho_0/(2\pi)$, so that $\cP_{k\geq 1}(x_w)=0$, we
can now relate $\cP_{k}$ to $\cP_{k-1}$ and then compute iteratively
the $C_k$'s.

In steady-state, the master equation gives for $x>x_w$, order by order
in $\bar \lambda$:
\begin{align}
  \label{eq:masterorder}
  0&=-\partial_x (v \cos\theta \cP_k-\frac{\mu_t D_r}{\kappa\mu_r}(x-x_w)\cP_{k-1})+D_r\partial_\theta^2\cP_k-D_r\partial_{\theta}(\sin(2\theta)\cP_{k-1}), \quad k\ge 1 \\
  0&=-\partial_x (v \cos\theta \cP_0)+D_r\partial_\theta^2\cP_0\;.
\end{align}
Multiplying Eq.~(\ref{eq:masterorder}) by an arbitrary function
$f(\theta)$ and integrating over $\theta$ and $x$, one gets
\begin{align}
  \label{eq:masterorder2}
  \int_{x_w}^\infty& dx\int_0^{2\pi} d\theta f''\cP_k=-\int_{x_w}^\infty dx\int_0^{2\pi} d\theta f'\sin(2\theta)\cP_{k-1}, \quad k\ge 1 \\
  \label{eq:masterorder3}
  \int_{x_w}^\infty&dx \int_0^{2\pi} d\theta f''\cP_0=-\frac{1}{D_r}\int_0^{2\pi} d\theta v \cos\theta f\cP_0(x_w,\theta)=-\frac{v\rho_0}{2\pi D_r}\int d\theta \cos\theta f\;.
\end{align}
For conciseness, we define the operators $T$ and $T^*$
\begin{equation}
  \label{eq:operators}
  T(f)=\sin(2\theta) \int d\theta f \quad,\qquad T^*(f)=\cos\theta \int d\theta\int d\theta f\;,
\end{equation}
where the integral signs refer to indefinite integrals, to rewrite
Eqs.~(\ref{eq:masterorder2}-\ref{eq:masterorder3}) as
\begin{align}
  \int_{x_w}^\infty& dx\int_0^{2\pi} d\theta g (\theta)\cP_k=-\int_0^\infty dx\int_0^{2\pi} d\theta T(g(\theta))\cP_{k-1}, \quad k\ge 1 \\
  \int_{x_w}^\infty&dx \int_0^{2\pi} d\theta g(\theta) \cP_0=-\frac{1}{D_r}\int_0^{2\pi} d\theta v T^*(g(\theta)) \cP_0(x_w,\theta)=-\frac{v\rho_0}{2\pi D_r}\int d\theta T^*(g(\theta))\;,
\end{align}
where $g=f''$. The $C_k$'s then reduce to the explicit integrals
\begin{equation}
  \label{eq:ckoperator}
  C_k=(-1)^{k+1}\frac{v\rho_0}{2\pi D_r}\int_0^{2\pi}d\theta T^*T^{k+1}(\cos\theta)\;,
\end{equation}
where we use $\sin\theta\sin(2\theta)=T(\cos\theta)$ so that
$T^k(\sin\theta\sin(2\theta))=T^{k+1}(\cos\theta)$.

Let us now compute the $C_k$'s. By inspection, one sees that $T^k(\cos\theta)$ is of the form
\begin{equation}
  \label{eq:defalphak}
  T^k(\cos\theta)=\sum_{i=0}^k \alpha_i^k \cos((2i+1)\theta) \;,
\end{equation}
where the coefficients $\alpha_i^k$ obey the recursion
\begin{align}
  \alpha_0^0&=1, \qquad \alpha_{j>0}^0=0\;, \\
  \alpha_0^{k+1}&=\frac{\alpha_0^k}{2}+\frac{\alpha_1^k}{6} \\
  \alpha_i^{k+1}&=\frac{1}{2}\left( \frac{\alpha_{i+1}^k}{2i+3}-\frac{\alpha_{i-1}^k}{2i-1}\right) \;,\\
  \alpha_{k}^{k+1}&=-\frac{1}{2}\frac{\alpha_{k-1}^k}{2k-1}\;,\\
  \alpha_{k+1}^{k+1}&=-\frac{1}{2}\frac{\alpha_k^k}{2k+1}\;,
\end{align}
which solution is
\begin{equation}
  \label{eq:alphas}
  \alpha_j^k=\frac{(-1)^j}{k+1}\frac{(2j+1)}{(k+j+1)!}\prod_{i=0}^j(k+1-i)\;.
\end{equation}
After the application of $T^*$ in Eq.~\eqref{eq:ckoperator}, the only term that
contributes to $C_k$ in $T^k(\cos\theta)$ is $\alpha_0^k=\frac{1}{(k+1)!}$,
because $\int d\theta \cos\theta\cos[(2i+1)\theta]=0$ for $i>0$. One
thus finally gets
\begin{align}
  C_k&=(-1)^{k}\frac{v\rho_0}{2\pi D_r}\int_0^{2\pi}d\theta \alpha_0^{k+1}\cos^2(\theta) 
  =(-1)^{k}\frac{v\rho_0}{2D_r(k+2)!}\;.
\end{align}
\subsubsection{Approximate expression for the pressure}
The series~\eqref{eq:pressureseries} can now be resummed to yield
\begin{equation}
  \label{eq:pressure_torque}
  P=\frac{v^2}{2\mu_t D_r}\rho_0\left(1-\sum_{k=0}^\infty(-1)^k\frac{\bar\lambda^{k+1}}{(k+2)!}\right) =P_I\frac{1-e^{-\bar\lambda}}{\bar\lambda}\;,
\end{equation}
where $P_I$ is the ideal gas pressure. As expected, the pressure tends
to $P_I$ as $\bar\lambda\to 0$.

As can be seen in the right panel of Fig.~1 in the main text, the
approximation that the wall does not affect the probability density
for $x\leq x_w$ is not satisfied when $\bar\lambda$ is large. However,
this happens only when $P(\bar \lambda)$ is already very small, so
that the analytic formula Eq.~(\ref{eq:pressure_torque}) compares very
well with the $P(\bar \lambda)$ curve obtained numerically, as shown
in Figure 1 of main text.

\section{Non--Boltzmann Distribution}

While the analytical computation of the full distribution for RTPs and
ABPs in two dimensions is beyond the scope of this paper, here we show
explicitly that the steady--state density is not a Boltzmann
distribution for 1D RTPs. The master equation for the probability
densities of right and left-movers ($\cP_+(x,t)$ and $\cP_-(x,t)$)
is given by (see Ref.~(26) of the main text)
\begin{eqnarray}
\partial_t \cP_+ = -\partial_x \left( v- \mu_t \partial_x V \right) -\frac{\alpha}{2}\left(\cP_+ - \cP_-\right)\;, \nonumber \\
\partial_t \cP_- = -\partial_x \left( -v- \mu_t \partial_x  V \right) -\frac{\alpha}{2}\left(\cP_- - \cP_+\right)\;.
\end{eqnarray}
Note that $D_t=0$ for this system. The equation for the steady-state
density then reads
\begin{equation}
\partial_x \left[ \left(v^2-\mu_t^2(\partial_x V)^2 \right)\rho \right]+\alpha \mu_t (\partial_x V) \rho=0\;.
\label{eq:1druntumble}
\end{equation}
First, rescale the potential so that the equation reduces to
\begin{equation}
\partial_x \left[ \left(1-(\partial_x \tilde V)^2 \right)\rho \right]+g (\partial_x \tilde V) \rho=0\;,
\end{equation}
with $g=\alpha/v$ and $\tilde V=V\mu_t/v $. The steady state
distribution is then given by
\begin{equation}
\rho(x)=\rho_0 e^{-Q}\;,
\end{equation}
and
\begin{equation}
Q=\ln[1-(\partial_x \tilde V(x))^2]
+\int_0^x dx' \frac{g\partial_{x'}\tilde V(x')}{1-(\partial_{x'}\tilde V(x'))^2} \;.
\end{equation}
The probability distribution is non-local inside the wall and not
given by a Boltzmann distribution. (Note that particles are confined
within the region $[0,x^*]$ where $(\partial_x \tilde V)^2<1$ and $\rho(x)=0$
outside.) 

Despite the absence of a Boltzmann distribution, the pressure is well
defined (as for the 2D case considered in the text). To see this
explicitly in one dimension consider the expression for the pressure
\begin{equation}
P=\frac{v}{\mu_t}\int_0^{x^*} \partial_x V(x) \rho(x)  \;,
\end{equation}
with $\partial_{x}V(x^*)=1$. Then using the explicit expression of the
steady-state distribution, $P$ can be written as
\begin{equation}
P=-\rho_0\frac{v}{g\mu_t}\int_0^{x^*}dx \partial_x e^{-g\int_0^x dx' \frac{\partial_{x'}V(x')}{1-(\partial_{x'}V(x'))^2} }\;,
\end{equation}
so that
\begin{equation}
P=-\rho_0\frac{v}{g\mu_t}\left(e^{-g\int_0^{x^*} dx' \frac{\partial_{x'}V(x')}{1-(\partial_{x'}V(x'))^2} }-1\right)\;.
\end{equation}
Now, since at the upper bound of the integral within the exponential
the integrand diverges we have
\begin{equation}
P=\rho_0\frac{v}{g \mu_t} =\rho_0 \frac{v^2}{\alpha \mu_t} \;.
\end{equation}

\section{Anisotropic pressure}

We consider spherical particles whose speeds depend on their
direction of motion $\theta$. As discussed in the main text, such
situations could arise, for example, when the motion takes place on a
corrugated surface. For simplicity, we consider only run--and--tumble
particles ($D_r=0$). The case of active Brownian particles can be
treated following the same argument.

In steady--state, the master equation yields
\begin{equation}
0=-\partial_x\left[ \left( v(\theta) \cos \theta -\mu_t \partial_x V-D_t \partial_x \right) \cP(\theta,{\bf x})\right]-\alpha \cP+\frac{\alpha}{2\pi}\int d\theta'\cP(\theta',{\bf x})\;.
\label{eq:2druntumbleBA}
\end{equation}
We want to restrict ourselves to cases where the bulk currents along
any direction vanish (the system is therefore uniform in the bulk),
which we achieve by assuming that $v(\theta+\pi)=v(\theta)$. Following
the same steps that lead to Eq.~(4) in the main text, we get in steady
state
\begin{align}
 0&=-\partial_x(\tilde m_1-\mu_t\rho\partial_xV -D_t\partial_x \rho) \;,\\
 0&=-\partial_x\left[\int_0^{2\pi} v(\theta)^2\cos^2(\theta)\cP d\theta-\mu_t\partial_xV \tilde m_1-D_t\partial_x \tilde m_1 \right]-\alpha \tilde m_1\;,
\label{eq:2dpartialB}
\end{align}
where we have defined $\tilde m_1=\int_0^{2\pi} v(\theta)\cos(\theta)\cP
d\theta$ (which differs from $m_1$ in section~\ref{sec:derivationP} because it
includes the speed).

From these two equations, we can express the mechanical pressure as a
function of the bulk density and $v(\theta)$, as
\begin{align}
  P=\int_0^x \rho(x)\partial_xV dx =\frac{1}{\mu_t}\int_0^x (\tilde m_1 -D_t\partial_x \rho)=\left(\frac{D_t}{\mu_t}+\frac{\int_0^{2\pi}d\theta v^2(\theta)\cos^2(\theta)}{2\pi\alpha\mu_t}\right)\rho_0\;.
\end{align}
This holds for a wall perpendicular to the $\hat x$ axis. For a wall
tilted by an angle $\phi$, one obtains the anisotropic pressure
\begin{align}
  P(\phi)=\left(\frac{D_t}{\mu_t}+\frac{\int_0^{2\pi}d\theta v^2(\theta)\cos^2(\theta-\phi)}{2\pi\alpha\mu_t}\right)\rho_0\;,
\end{align}
which is Eq.~(9) of the main text.

\section{Interacting active brownian particles}

In the following we study ABPs with aligning interactions
(Section~\ref{sec:aligning}) and quorum-sensing interactions
(Section~\ref{sec:quorum}). In particular, we derive exact expressions
for the pressure $P$ in terms of microscopic correlators {evaluated} near the
wall. These show $P$ to depend explicitly on the details of the
interaction with the wall, hence forbidding the existence of equations
of state.

\subsection{Aligning particles}
\label{sec:aligning}
We consider a system of $N$ spherical ABPs which can exert torque on
each other, for instance to promote the alignment of their directions
of motion, {but which do not feel any wall-torque}. The positions and {orientations} of the particles evolve
according to the It\=o-Langevin equations
\begin{align}
  \frac{d\vec r_i}{dt}&=\vec v-\mu_t\partial_x V+\sqrt{2 D_t} \vec \eta_i(t) \\
  \frac{d\theta_i}{dt}&=\mu_r\sum_j F(\theta_j-\theta_i,\r_i,\r_j)+\sqrt{2 D_r} \xi_i(t)
\end{align}
where $\vec \eta_i$ and $\xi_i$ are uncorrelated Gaussian white noises
of unit variance and appropriate
dimensionality. $F(\theta_j-\theta_i,\r_i,\r_j)$ is the torque exerted
by particle $j$ on particle $i$.

{We now} define a microscopic density field $\cP(\r,\theta)$ as
\begin{equation}
  \label{eq:def-P}
  \cP(\vec r,\theta)=\sum_{i=1}^N \delta(\vec r-\vec r_i)\delta(\theta-\theta_i)
\end{equation}
Following~\cite{Farrell}, its evolution equation is given by
\begin{align}
  \label{eq:master-aligning}
  \partial_t \cP(\vec r,\theta) = - \nabla \cdot& [({\bf v} - \mu_t \nabla V(x)) \cP(\vec r,\theta)-D_t \nabla \cP(\vec r,\theta)] +\nabla \cdot\left(\sqrt{2 D_t \cP}\vec\eta\right)+\partial_\theta \left(\sqrt{2 D_r \cP}\vec\xi\right)\nonumber\\
&-\partial_\theta\left[\mu_r \int dr'\int_0^{2\pi} d\theta'F(\theta'-\theta,\vec r,\vec r') \cP(\vec r,\theta)\cP(\vec r',\theta')\right]+ D_r \partial_{\theta}^2\cP(\vec r,\theta)
\end{align}
where the integral $\int d\vec r'$ is performed over all space.

We then follow the same reasoning as for non-interacting particles to
derive an expression for the pressure. We first average
Eq.~\eqref{eq:master-aligning} in steady-state, assuming translational
invariance along $y$, to get
\begin{multline}
  \label{eq:master-aligning2}
  0 = - \partial_x [({\bf v} - \mu_t \partial_x V(x)) \langle\cP\rangle-D_t \partial_x \langle\cP\rangle]
-\partial_\theta\left[\mu_r\int dr'\int_0^{2\pi} d\theta'F(\theta'-\theta,\vec r,\vec r') \langle\cP(\vec r,\theta)\cP(\vec r',\theta')\rangle\right]\\+ D_r \partial_{\theta}^2\langle\cP\rangle
\end{multline}
where the brackets $\langle \cdot \rangle$ denote averaging over noise
realisations. Note that the noise terms average {to zero} due to {our use of} the It\=o
convention.

Multiplying Eq.~(\ref{eq:master-aligning2}) by $1$ and $\cos\theta$
and then integrating over $\theta$ gives the analog of Eq.~\eqref{eq:ss-rho} and~\eqref{eq:ss-m}

\begin{align}
  0&= - \partial_x [v m_1 - \mu_t\rho (\partial_x V) -D_t \partial_x \rho] \label{rho-aligning}\\
  D_r m_1&=-\partial_x \left[v\frac{\rho+m_2}{2} -\mu_t  m_1 (\partial_xV)-D_t \partial_x m_1\right]\nonumber \\  & \qquad\qquad-\mu_r\int_0^{2\pi}\!\!\sin\theta\int dr'\int_0^{2\pi}d\theta'F(\theta'-\theta,\vec r,\vec r') \langle\cP(\vec r,\theta)\cP(\vec r',\theta')\rangle \label{m-aligning}
\end{align}
where $m_n(x)=\int_0^{2\pi} \cos(n \theta)\langle\cP(x,\theta)\rangle d\theta$ and $\rho(x)=\int_0^{2\pi} \langle\cP(x,\theta)\rangle d\theta$.

Inserting Eq.~(\ref{m-aligning}) in Eq.~(\ref{rho-aligning}) allows us
to rewrite the pressure $P=\int_0^\infty dx
\rho \partial_x V$ {exactly} as
\begin{equation}
  \label{eq:pressure-aligning}
  P = \left[\frac{v^2}{2\mu_tD_r}+\frac{D_t}{\mu_t}\right] \rho_0 - \frac{ v\mu_r}{\mu_tD_r}\int_{0}^\infty  \!\!\! dx\int_{-\infty}^\infty  \!\!\! dy \int_0^{2\pi} \!\!\! d \theta\int  \! d\vec r' \int_0^{2\pi} \!\!\! d \theta' 
  F(\theta'-\theta,\vec r,\vec r') \sin\theta \langle \cP(\vec r,\theta)\cP(\vec r',\theta')\rangle 
\end{equation}
We see that, {just as in} Eq.~(4) in main text, the mechanical pressure
depends explicitly on the density {${\cal P}({\bf r},\theta)$} close to the wall,
which in turn depends on the detail of the interaction $V(x)$ between
the particles and the wall. There is thus no equation of state.

Using the microscopic definition of $\cP$, Eq.~(\ref{eq:def-P}), one
can rewrite the integral in Eq.~(\ref{eq:pressure-aligning}) as a sum
over all particles, more suitable to numerical measurements:
\begin{equation}
  \label{eq:pressure-micro-aligning}
  P = \left[\frac{v^2}{2\mu_tD_r}+\frac{D_t}{\mu_t}\right] \rho_0 - \frac{ v\mu_r}{\mu_tD_r} \langle \sum_{i,j=1}^N F(\theta_j-\theta_i,\vec r_i,\vec r_j) \sin\theta_i \Theta(x_i) \rangle
\end{equation}
where $\Theta(x_i)=1$ if $x_i>0$ and zero otherwise. In
Fig.~\ref{fig:pressurealigningtheory}, we compare measurements of $P$
from the force applied on the confining wall and from
Eq.~(\ref{eq:pressure-micro-aligning}), for a particular choice of
$F$. They show perfect agreement, thus {confirming 
Eq.~(\ref{eq:pressure-aligning})}.

\begin{figure}
  \includegraphics[width=.33\textwidth]{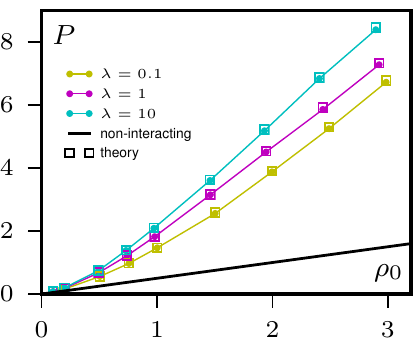}
  \caption{{Lack of equation of state for {ABPs with interparticle alignment interactions but no wall torques}. The mechanical
    force per unit area $P$ exerted on the wall is equal to its
    theoretical expression~\eqref{eq:pressure-micro-aligning} and
    depends on the stiffness $\lambda$ of the wall potential. The
    torque exerted by particle $j$ on particle $i$ is
    $F(\theta_j-\theta_i,\r_i,\r_j)=\frac{\gamma}{\mathcal{N}(\r_i)}\sin(\theta_j-\theta_i)$
    if $|\r_j-\r_i|<R$ and $0$ otherwise, where $\mathcal{N}(\r_i)$ is
    the number of particles interacting with particle $i$. ($v=1$,
    $D_r=1$, $D_t=0$ and $\gamma=2$.)}}
  \label{fig:pressurealigningtheory}
\end{figure}

\subsection{Quorum-sensing interactions}
\label{sec:quorum}
A similar path can be followed to compute the pressure exerted by ABPs
that adapt their swim speed to the local density computed through a
coarse-graining kernel $\bar\rho(\vec r)=\sum_i K(|\vec r-\vec r_i|)$,
where the sum runs over all particles. The dynamics of the system is
now given by the It\=o-Langevin equations
\begin{align}
  \frac{d\vec r_i}{dt}&=v(\bar \rho)\vec e_i-\mu_t\partial_x V+\sqrt{2 D_t} \vec \eta_i(t) \\
  \frac{d\theta_i}{dt}&=\sqrt{2 D_r} \xi_i(t)
\end{align}

As before, the dynamics of the density field can be obtained using It\=o {calculus}~\cite{Farrell}
\begin{align}
  \label{eq:master-vrho}
  \partial_t \cP(\vec r,\theta) = - \nabla \cdot [(v(\bar\rho)\vec e_\theta & - \mu_t \nabla V(x)) \cP(\vec r,\theta)-D_t \nabla \cP(\vec r,\theta)]  + D_r \partial_{\theta}^2\cP(\vec r,\theta)\\
&+\nabla \cdot\left(\sqrt{2 D_t \cP}\vec\eta\right)+\partial_\theta \left(\sqrt{2 D_r \cP}\vec\xi\right)\nonumber
\end{align}

By the same procedure as for aligning particles (except that we first
multiply Eq.~\eqref{eq:master-vrho} by $v(\bar \rho)$ for the second
equation) we get the two relations
\begin{align}
  0&= - \partial_x [\langle v(\bar\rho) \hat m_1\rangle - \mu_t\rho (\partial_x V) -D_t \partial_x \rho] \\
  D_r \langle v(\bar\rho)\hat m_1\rangle &=-\left\langle v(\bar \rho)\partial_x \left[v(\bar \rho)\frac{\hat\rho+\hat m_2}{2} -\mu_t \hat m_1 (\partial_xV)-D_t \partial_x \hat m_1\right]\right\rangle
\end{align}
where $\hat m_n(x)=\int_0^{2\pi} \cos(n \theta)\cP(x,\theta) d\theta$
and $\hat \rho(x)=\int_0^{2\pi} \cP(x,\theta) d\theta$ {are 
fluctuating} quantities whose averages are $m_n$ and $\rho$.

We can now rewrite the pressure using these two equalities:
\begin{align}
  \label{eq:pressure-vrho}
  P =& \int_0^\infty dx \rho \partial_x V =\frac{1}{\mu_t}\int_0^\infty dx \left[\langle v(\bar\rho) \hat m_1\rangle-D_t \partial_x \rho\right] \\
  =&\frac{D_t}{\mu_t} \rho_0-\frac{1}{D_r\mu_t}\int_0^\infty dx \left\langle v(\bar \rho)\partial_x \left[v(\bar \rho)\frac{\hat\rho+\hat m_2}{2} -\mu_t \hat m_1 (\partial_xV)-D_t \partial_x \hat m_1\right]\right\rangle
\end{align}
Integrating by part the last integral, we obtain
\begin{align}
  \label{eq:pressure-vrho2}
  P=&\frac{\langle v(\bar \rho)^2(\hat\rho+\hat m_2)\rangle_0}{2\mu_tD_r}-\frac{D_t\langle v(\bar \rho)\partial_x\hat m_1)\rangle_0}{\mu_tD_r} +\frac{D_t}{\mu_t} \rho_0\\\nonumber
  &+\frac{1}{D_r\mu_t}\int_0^\infty dx \left\langle \partial_x v(\bar \rho) \left[v(\bar \rho)\frac{\hat\rho+\hat m_2}{2} -\mu_t \hat m_1 (\partial_xV)-D_t \partial_x \hat m_1\right]\right\rangle
\end{align}
where the brackets $\langle \cdot \rangle_0$ denote an average done in
the bulk of the system.

As for aligning particles, one can use Eq.~(\ref{eq:def-P}) to obtain
a ``microscopic expression'' for $P$ which is more suitable for
numerical {evaluation}:
\begin{align}
  \label{eq:pressure-micro-vrho}
  P=&\frac{D_t}{\mu_t} \rho_0+\sum_{i=1}^N \left( \frac{\langle v(\bar \rho_i)^2(1+\cos(2\theta_i))\rangle_0}{2\mu_tD_r}+\frac{2 D_t\langle \partial_{x_i} v(\bar \rho_i) \cos\theta_i\rangle_0}{\mu_tD_r}\right)\\
  +&\sum_{i=1}^N \Theta(x_i) \frac{1}{D_r\mu_t} \left\langle \partial_{x_i} v(\bar \rho_i) \left[v(\bar \rho_i)\frac{1+\cos(2\theta_i)}{2} -\mu_t \cos\theta_i (\partial_xV)\right] 
+   D_t(\partial_{x_i}^2 v(\bar \rho_i)) \cos\theta_i \right\rangle\nonumber
\end{align}
{Here, for ease of notation, we have written} $\bar\rho_i=\bar \rho(\vec r_i)$. Again
this exact formula shows that no equation of state relates the
mechanical pressure to bulk properties of the system.

\bibliographystyle{Science}

\end{document}